\documentclass[final]{siamltex}
\usepackage{graphicx}

\title{Deflated Iterative Methods for Linear Equations with Multiple
Right-Hand Sides\footnotemark[1]}

\author{Ronald B. Morgan\footnotemark[2]
\and Walter Wilcox\footnotemark[3] }

\begin{document}
\bibliographystyle{plain}

\maketitle

\renewcommand{\thefootnote}{\fnsymbol{footnote}}
\footnotetext[1]{This work was partially supported by the National Science Foundation, Computational Mathematics Program under grant 0310573 and the National Computational Science Alliance.  It utilized the SGI Origin 2000 and IBM p690 systems at the University of Illinois.  The first author was also supported by the Baylor University Sabbatical Program.}
\footnotetext[2]{Department of Mathematics, Baylor
University, Waco, TX 76798-7328 ({\tt Ronald\_Morgan@baylor.edu}).}
\footnotetext[3]{Department of Physics, Baylor
University, Waco, TX 76798-7316.  ({\tt Walter\_Wilcox@baylor.edu}).}
\renewcommand{\thefootnote}{\arabic{footnote}}

\begin{abstract}
A new approach is discussed for solving large nonsymmetric systems of linear equations with multiple right-hand sides.  The first system is
solved with a deflated GMRES method that generates eigenvector information
at the same time that the linear equations are solved.  Subsequent systems
are solved by combining restarted GMRES with a projection over the previously determined eigenvectors.  This approach offers an alternative to block methods, and it can also be combined with a block method.  It is useful when there are a limited number of small eigenvalues that slow the convergence.  An example is given showing significant improvement for a problem from quantum chromodynamics.  The second and subsequent right-hand sides are solved much quicker than without the deflation.  This new approach is relatively simple to implement and is very efficient compared to other deflation methods.
\end{abstract}



\pagestyle{myheadings}
\thispagestyle{plain}
\markboth{R. B. MORGAN and W. WILCOX}{DEFLATION FOR MULTIPLE RIGHT-HAND SIDES}

\section{Introduction}

Large systems of linear equations $Ax=b$ arise in many areas of science.  Often
there are many right-hand sides associated with a single matrix.  It
is then important to consider these systems together and take advantage of the
relationship between them.  Here we consider solving the systems with iterative methods, and we assume the matrix is real nonsymmetric or complex non-Hermitian.

A standard way of dealing with multiple right-hand sides is to use a
block method (see for example~\cite{OL80,Sa96,FrMa,SiGa96B,ChSa}).  Krylov subspaces are generated with each right-hand side as starting vector and are used together.  Alternatives are presented in~\cite{SiGa,SiGa96}, including a method with the right-hand sides solved individually using Richardson iteration with a polynomial generated from GMRES~\cite{SaSc} applied to the first right-hand side.  Related methods that could be applied to multiple right-hand sides are given in \cite{NaReTr, StVa, CaRe}.  See~\cite{PadeStMaJoMa} for a method for multiple right-hand sides that can also handle a changing matrix.  This is more complicated than the approaches that will be given in this paper but is related in that the method GMRES-DR is used.  A seed method for nonsymmetric problems with multiple right-hand sides is given in \cite{KiMiRa}.  

Here we present a new option.  Eigenvector information generated while solving the first right-hand side is used to help solve the other right-hand sides.  This approach can be helpful for difficult problems with small eigenvalues.

The first right-hand side is solved with a deflated GMRES method, which also generates approximations to eigenvectors.  These approximate eigenvectors can then be used to deflate eigenvalues from the solution of the linear equations with the subsequent right-hand sides.  A fairly simple approach yields a useful method.  Specifically, we alternate cycles of regular GMRES with projections over the approximate eigenvectors.   For situations when a block approach is particularly desirable, it is possible to use a deflated block method both for the initial phase in which the eigenvectors are generated and for solution of subsequent right-hand sides.

Section 2 reviews deflated GMRES methods and a projection that will be used. 
Section 3 gives the projected version of GMRES and gives examples comparing it to other methods.  Section 4 has a block approach.  Section 5 discusses application to lattice quantum chromodynamics (QCD).  It is shown that the method can be very effective for an important problem.

\section{Deflated GMRES and Projections}

Small subspaces for restarted GMRES can slow convergence for difficult problems.  Deflated versions of restarted
GMRES~\cite{GMRES-E,KhYe,ErBuPo,ChSa,Sa95B,BaCaGoRe,BuEr,LCMo,DS99,GMRES-IR,
GMRES-DR} can improve this, when the problem is difficult due to a few
small eigenvalues.  One of these approaches is related to Sorensen's
implicitly restarted Arnoldi method for eigenvalues~\cite{So} and is called
GMRES with implicit restarting~\cite{GMRES-IR}.  A mathematically
equivalent method, called GMRES with deflated restarting
(GMRES-DR)~\cite{GMRES-DR}, is also related to Wu and Simon's restarted
Arnoldi eigenvalue method~\cite{WuSi}.  See~\cite{Arnoldi-R,St01,HRAM} for
some other related eigenvalue methods.  

We will concentrate on GMRES-DR, because it is efficient and relatively
simple.  Approximate eigenvectors corresponding to the small eigenvalues are
computed at the end of each cycle and are put at the beginning of the next
subspace.  Letting $r_0$ be the initial residual for the linear equations
at the start of the new cycle and $\tilde y_1, \ldots
\tilde y_k$ be harmonic Ritz vectors~\cite{IE,Fr92,PaPavdV,IEN}, the
subspace of dimension $m$ used for the new cycle of GMRES-DR(m,k) is
\begin{equation}
Span\{\tilde y_1, \tilde y_2, \ldots \tilde y_k, r_0, A r_0, A^2 r_0, A^3
r_0, \ldots ,A^{m-k-1} r_0 \}. \label{ss}
\end{equation}
This can be viewed as a Krylov subspace generated with starting vector
$r_0$ augmented with approximate eigenvectors.  Remarkably, the whole
subspace turns out to be a Krylov subspace itself (though not with $r_0$ as
starting vector)~\cite{GMRES-IR}.
Once the approximate eigenvectors are moderately accurate, their inclusion
in the subspace for GMRES essentially deflates the corresponding
eigenvalues from the linear equations problem.  

The approximate eigenvectors in GMRES-DR span a small Krylov subspace and so are generated in a compact form 
\begin{equation}
 AV_k = V_{k+1} \bar H_k, \label{recur1}
\end{equation} 
where $V_k$ is a $n$ by $k$ orthonormal matrix, $V_{k+1}$ is the same except for an extra column, and $\bar H_k$ is a full $k+1$ by $k$ matrix.  The columns of $V_k$ span a Krylov subspace, and more importantly, they span the subspace of approximate eigenvectors.  At the end of a cycle of GMRES-DR, the harmonic Ritz values are computed.  The matrix $V_k$ is then formed so that it is orthonormal and has columns spanning the harmonic Ritz vectors corresponding to desired harmonic Ritz values.  Complex harmonic Ritz vectors are split into real and imaginary parts so that complex arithmetic is not needed.  See~\cite{GMRES-DR} for the GMRES-DR algorithm that generates $V_{k+1}$ and $H_k$.  Note this compact form is similar to an Arnoldi recurrence, and it allows access to both the approximate eigenvectors and their products with $A$ while requiring storage of only $k+1$ vectors of length $n$.    

Next we review minimum residual (minres) projections, and then give the specific minres projection which will be needed.  See Saad~\cite{Sa96} for more on the minres projection and other projections.  

\vspace{.10in}
\begin{center}
\textbf{Minres Projection}
\end{center}
\begin{enumerate}
 \item Let the current approximate solution be $x_0$ and the current system
of equations be $A(x-x_0) = r_0$.  Let $V$ be a matrix (preferably orthonormal) with columns spanning the desired projection subspace.  
 \item Solve $V^T A^T A Vd = V^T A^T r_0$.
 \item The new approximate solution is $x_{new} = x_0 + Vd$.
 \item The new residual vector is $r_{new} = r_0 - AVd $.
\end{enumerate} 
\vspace{.15in}

Next we want the projection to be over the subspace spanned by the columns of the matrix $V_k$ from Equation~(\ref{recur1}).  The solution of the system of equations in step 2 becomes a least squares problem (as in the derivation of GMRES~\cite{Sa96}).

\vspace{.10in}
\begin{center}
\textbf{Minres Projection for $V_k$}
\end{center}
\begin{enumerate}
 \item Let the current approximate solution be $x_0$ and the current system
of equations be $A(x-x_0) = r_0$.  Let $V_{k+1}$ and $\bar H_k$ be the matrices from Equation (\ref{recur1}).  
 \item Solve min$||c - \bar H_k d||$, where $c = V_{k+1}^T r_0$.
 \item The new approximate solution is $x_k = x_0 + V_{k}d$.
 \item The new residual vector is $r_k = r_0 - AV_{k} d = r_0 - V_{k+1} \bar H_k d$.
\end{enumerate} 
\vspace{.15in}

This projection is fairly inexpensive, requiring only $3k+2$ vector operations (dot products and daxpys) of length $n$.

\section{Deflated GMRES for Multiple Right-hand Sides}

If a method such as GMRES-DR is used for the first right-hand side, eigenvector information is generated while the linear equations are solved.  We wish to use this information to assist with the solution of the other right-hand sides.  We will suggest three ways of doing this.  The main focus will be on the third approach, and it will be compared against the first two.

One possible way to use the approximate eigenvectors is to put them into the subspaces used for GMRES.  Such a method is called GMRES-E in~\cite{GMRES-E}.  The subspace has a basis like~(\ref{ss}) for GMRES-DR, but with the approximate eigenvectors going last in forming the basis. For subsequent right-hand sides the eigenvectors are already computed, so they can be left fixed.  

Another approach to deflating eigenvalues for the subsequent right-hand sides is to use the approximate eigenvectors to build a preconditioner for GMRES.  Burrage and Erhel propose a method called DEFLATION~\cite{BuEr}.  They do not consider multiple right-hand sides, but their method can be adapted by using the preconditioner from DEFLATION, but not the portion of DEFLATION that computes eigenvectors.  For both these approaches to deflation, there are significant costs compared to simple restarted GMRES.  With GMRES-E, there are $k$ additional vectors that are orthonormalized.  With DEFLATION, every iteration requires additional work in applying the preconditioner.  The approach we discuss next is more efficient. 

A relatively simple way of deflating is to use the projection mentioned in the previous section.  Projections over the subspace of approximate eigenvectors can be alternated with cycles of GMRES.  A major difference between this approach and  those mentioned in the last two paragraphs is that the eigenvectors are not needed during the GMRES iteration.  This approach can be much cheaper if many eigenvectors are used.  We call this method GMRES(m)-Proj(k), where $m$ is the dimension of the Krylov subspaces used in the GMRES cycles and $k$ is the number of approximate eigenvectors.  These projections are mentioned briefly in~\cite{GMRES-DR} for the case of just one right-hand side.  For multiple right-hand sides, some preliminary experiments are reported in~\cite{Qcdconf} for lattice QCD problems.  Further QCD experiments are at the end of this section.  

The GMRES-Proj method that follows is for all right-hand sides except for the first one.  Superscripts identify which right-hand side the vectors are associated with.  
\vspace{.10in}
\begin{center}
\textbf{GMRES(m)-Proj(k)}
\end{center}
\begin{enumerate}
 \item After applying the initial guess $x_0^{(i)}$, let the system
of equations be $A(x^{(i)}-x_0^{(i)}) = r_0^{(i)}$.  
 \item If it is known that the right-hand sides are related, project over the previous computed solution vectors.
 \item Apply the Minres Projection for $V_k$.  This uses the $V_{k+1}$ and $\bar H_k$ matrices developed while solving the first right-hand side with GMRES-DR.
 \item Apply one cycle of GMRES(m).
 \item Test the residual norm for convergence (can also test during the GMRES cycles).  If not satisfied, go back to step 3.
\end{enumerate} 
\vspace{.15in}

For step 2, we use a minres projection (see the previous section) over each solution vector separately.  Section 3.7 has an example.

The projection in step 3 adds little to the cost of the method.  One cycle of GMRES requires about $m^2 + 2m$ length $n$ vector operations plus the cost of $m$ matrix-vector products and applications of the preconditioner.  The projection step uses just over $3k$ vector ops and requires no matrix-vector products.  If for example $m=15$, $k=10$, the matrix has five nonzeros per row and no preconditioning is used, then the projection adds only $10 \%$ additional cost to a cycle.  

\subsection{Experiment}

The first example uses a simple test matrix for which deflation is important, because there are some small eigenvalues.  

{\it Example 1.}
The matrix is of size $n=2000$ and is bidiagonal with $0.1, 1, 2, 3, \ldots,$ $1998, 1999$ on the main diagonal and $1$'s on the superdiagonal.  GMRES-DR(25,10) has subspaces of total dimension 25 including 10 approximate eigenvectors.  It is applied to a randomly generated first right-hand side (random with unit normal distribution) until the residual norm has improved by a factor of $rtol = 10^{-6}$.  This takes 280 matrix-vector products. GMRES(15)-Proj(10) is then applied to a random second right-hand side.  We compare with several other methods, BiCGStab, Full GMRES, GMRES-DR(25,10) and GMRES(15).  The results for this second right-hand side are given in Figure 3.1.  Notice that GMRES-Proj has a big advantage over the other methods, because it deflates eigenvalues from the beginning.  The methods Full-GMRES, BiCGStab, and GMRES-DR must generate approximate eigenvectors as they proceed.  GMRES(15) restarts before it can develop effective approximate eigenvectors.  

We consider both matrix-vector products and flops, so that two cases can be simulated with this one test matrix.  For problems with expensive matrix-vector product or preconditioner, the matrix-vector product count matters.  For very sparse matrices without preconditioner, the flop count for this sparse test matrix is more relevant.  Of course many problems fall in between these extreme cases.

\begin{figure}
\vspace{.10in}
\begin{center}
\includegraphics[scale=0.5]{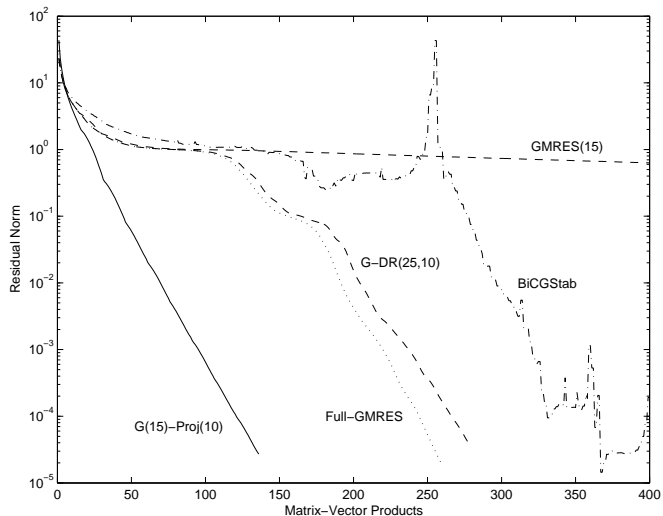}
\end{center}
\vspace{.10in}
\caption{Solution of second right-hand side.}
\end{figure}

For the first right-hand side, BiCGStab uses more matrix-vector products than GMRES-DR but considerably less flops.  BiCGStab needs 17.5 million flops versus 81.1 million for GMRES-DR to improve the residual norm by $10^{-6}$.  However, GMRES-Proj saves on both matrix-vector products and flops for the second right-hand side.  It uses 130 matrix-vector products compared to 365 for BiCGStab and 14.1 million flops versus 17.5.  Of course GMRES-Proj needs GMRES-DR applied first, but if there are a number of right-hand sides, the GMRES approach can still be competitive in terms of flops even for such a very sparse matrix.  For example, if there are 10 right-hand sides, then GMRES-DR for the first right-hand side and GMRES-Proj for the next nine takes 1405 matrix-vector products and 204.5 million flops.  BiCGStab on all 10 right-hand sides uses 4113 matrix-vector products and 197.6 million flops. 

\subsection{Effect in GMRES-Proj of the size of the GMRES subspace} 

We experiment with changing the size of the GMRES subspaces used to solve the subsequent right-hand sides.  The same problem from the experiment in the previous subsection is considered with again the first right-hand side solved with GMRES-DR(25,10).  This time ten right-hand sides are solved, with different $m$ values for GMRES(m)-Proj(10).  The first column of Table 3.1 gives the total number of matrix-vector products required to solve all ten systems.  We see that even with deflation, larger subspaces are helpful.  However, if the matrix-vector product is inexpensive, using a small value of $m$ such as $m=10$ might be more efficient than $m=20$ in spite of increased iterations.

\begin{table}
\caption{Changing $m$ and the frequency of projection for GMRES-Proj} 

\begin{center} \footnotesize
\begin{tabular}{|c|c|c|c|c|}  \hline\hline
       & project every cycle   & project every 5th  & project every 10th & project at 10,20, ... \\ \hline
$m$    & mat-vec's  & mat-vec's & mat-vec's & mat-vec's \\  
\hline \hline
5      & 2440    & 2577 & 2596 & 1962 \\ \hline
10     & 1658    & 1678 & 1636 & 1604 \\ \hline
15     & 1405    & 1411 & 1673 & 1910 \\ \hline
20     & 1298    & 1330 & 2107 & 2438 \\ \hline 
25     & 1257    & 1478 & 2521 & 2843 \\ \hline

\hline\hline 

\end{tabular} 
\end{center} 
\end{table}

\subsection{Projecting less frequently}

Table 3.1 also shows the effect of not projecting between every cycle of GMRES(m). For small values of $m$, it is not necessary to project very often.  Projecting reduces components of $r$ in the directions of the  eigenvectors corresponding to small eigenvalues, and these components may not need to be reduced futher for a while (basically until the rest of the residual vector has been reduced to the point that these components are again significant).  For $m=10$, projecting every tenth cycle is good enough.  For $m=20$, projecting every fifth is almost as good as projecting every cycle, while projecting every tenth is not as effective.  

In the case of $m=5$ and project every tenth cycle, the projections 
are performed before the 1st, 11th, 21st, \ldots cycles of GMRES.  If instead we project before the 10th, 20th, \ldots cycles, the results are better, with only 1962 matrix-vector products needed instead of 2596 (see the last column of Table 3.1).  It is not clear why the convergence is better if no projection is performed until after nine cycles of GMRES.   

\subsection{Solving the first right-hand side to greater accuracy}

We test here the notion that the eigenvector approximations provided by GMRES-DR might not be optimal at the same point that the linear equations are considered solved.  For the case of Example 1, solving 10 right-hand sides with $rtol$ of $10^{-6}$ requires 1405 iterations.  But if the system with the first right-hand side is solved to greater accuracy of $10^{-8}$, while the final nine systems are again are solved to $10^{-6}$, the total number of iterations drops to 1343.  This is in spite of the fact that solution of the first system takes 69 more iterations.  The average savings per subsequent right-hand side is 13.6 iterations.  However, solving the first right-hand side to even greater accuracy does not pay off.  With relative tolerance of $10^{-10}$, 1388 iterations are needed for all ten right-hand sides (exempting the first, the number actually stays the same as for the first $rtol$ being $10^{-8}$).

\subsection{Comparison with other deflation approaches}

While the GMRES-Proj method deflates eigenvalues with a projection that is separate from the GMRES phase, there are other ways of deflating eigenvalues as discussed at the beginning of this section.  We will compare GMRES-Proj with the versions of GMRES-E and DEFLATION that use eigenvectors to augment or precondition, but are adapted so they do not attempt to improve on the eigenvectors.  Note GMRES-DR is used on the first right-hand side to compute the eigenvectors for each method, then nine additional right-hand sides are solved.  We see from the results in Table 3.2 that the methods perform similarly.  However, as mentioned earlier, there is a difference in expense, since GMRES-Proj uses eigenvectors only once per cycle.  DEFLATION applies eigenvectors at every iteration, and the cost above the normal GMRES expense is $2k$ length $n$ vector operations per iteration or about $2km$ per cycle.  Costs for GMRES-E with $k$ approximate eigenvectors augmenting a $m$-dimensional Krylov subspace are a little greater than for DEFLATION (extra expense of about $2km + k^2$ length $n$ vector operations per cycle).   Meanwhile, as mentioned, GMRES-Proj requires about $3k$ extra per cycle.  So GMRES-Proj can be more efficient.  However, for very expensive matrix-vector product or for small $k$, GMRES-Proj may not be a significantly better way of deflating.

\begin{table}
\caption{Comparson of different deflation approaches} 

\begin{center} \footnotesize
\begin{tabular}{|c|c|c|c|}  \hline\hline
       & eigenvectors used for projection: &  eigenvectors in subspace: & eigenvector preconditioner: \\ 
$m$    & GMRES-Proj  &  GMRES-E            & DEFLATION  \\  
\hline \hline
5      & 2440    & 2565 & 2528 \\ \hline
10     & 1658    & 1658 & 1613 \\ \hline
15     & 1405    & 1405 & 1387 \\ \hline
20     & 1298    & 1296 & 1284 \\ \hline 
25     & 1257    & 1251 & 1241 \\ \hline

\hline\hline 

\end{tabular} 
\end{center} 
\end{table} 

\subsection{Comparison with block-QMR}

Here we show that the GMRES-Proj method can be competitive with block methods.  Specifically, we compare against block-QMR~\cite{FrMa} for 10 right-hand sides.  GMRES(15)-Proj(10) uses projections every fifth GMRES cycle.  Table 3.3 has the results for both the number of matrix-vector products and the number of flops as counted by MATLAB.  GMRES-Proj is better than block-QMR in terms of flops, since block-QMR with blocksize 10 has significant orthogonalization expense.  (Table 4.1 includes comparison with 40 right-hand sides.)

\begin{table}
\caption{Comparison with nonrestarted methods including block QMR} 

\begin{center} \footnotesize
\begin{tabular}{|c|c|c|c|}  \hline\hline
       & matrix-vector products  &  flops     \\  
\hline \hline
GMRES-DR + GMRES-Proj  & 1411    & 198.3   \\ \hline
Block-QMR              & 1782    & 567.5   \\ \hline
QMR, 10 times          & 5220    & 245.7   \\ \hline
BiCGStab, 10 times     & 4113    & 197.6   \\ \hline 

\hline\hline 

\end{tabular} 
\end{center} 
\end{table} 

\subsection{The case of related right-hand sides}

One expects intuitively that if the right-hand sides are closely related to each other then there should be a way to take advantage of the situation.  However, it is discussed in~\cite{bgdr} that block methods may not be successful at this.  We suggest here a simple way for GMRES-Proj to deal with this case.  For the second and subsequent right-hand sides, minres projections are done over all previously computed solutions (step 2 of the GMRES-Proj algorithm).  We project over each solution vector individually, but another option is to project over all at once~\cite{Fi98}.  

We again compare GMRES-Proj with block-QMR for 10 right-hand sides.  This time the first has random normal entries and all the others are equal to the first one plus $10^{-4}$ times a random vector.  GMRES-Proj is better able to take advantage of the related right-hand sides, because it solves them sequentially, and the results of one solution are available for the next problem.  GMRES-Proj uses only 521 matrix-vector products compared to 1702 for block-QMR.  In terms of flops, GMRES-Proj needs 110 million versus to 542 million for block-QMR.

\section{A Deflated Block Method}

Block methods are well-known for solving systems of equations with multiple right-hand sides.  We saw in Subsection 3.6 an example for which GMRES-Proj is better than block-QMR in terms of both matrix-vector products and flops.  However, with more right-hand sides in the next example, block-QMR is best in terms of matrix-vector products.  So there are situations with expensive matrix-vector product where block methods are needed.  This is particularly the case when the matrix-vector product can be efficiently applied to several right-hand sides simultaneously.  In this section we look at combining GMRES-Proj with block methods. 

A block GMRES method with deflated restarting called block-GMRES-DR is proposed in~\cite{bgdr}.  Here we look at the situation where a block method is worth considering, but there are more right-hand sides than can be efficiently solved with a single block run.  This could be either because the orthogonalization expense or storage would be too great or because not all right-hand sides are available at once.  We propose to solve the first group of say $p$ right-hand sides with block-GMRES-DR(m,p,k) (subspaces of dimension $m$ are used, the block-size is $p$, and $k$ approximate eigenvectors are generated).  The eigenvectors satisfy a block Arnoldi-like recurrence of the form $AV_k = V_{k+p}\bar H_m$, where $V_{k}$ is an orthonormal matrix with columns spanning the space of approximate eigenvectors, $V_{k+p}$ has $p$ columns appended to $V_k$, and $\bar H_k$ is $k+p$ by $k$.  For the next group of $p$ right-hand sides, we alternate minres projections over the approximate eigenvectors with cycles of block-GMRES.  Other right-hand sides are dealt with the same way, $p$ at a time.

\vspace{.10in}
\begin{center}
\textbf{Bl-GMRES(m,p)-Proj(k)}
\end{center}
\begin{enumerate}
 \item Apply initial guesses to the current $p$ right-hand sides being considered. 
 \item If it is known that the right-hand sides are related, project over the previous computed solution vectors.
 \item Apply the Minres Projection to all $p$ systems using the $V_{k+p}$ and $\bar H_k$ matrices developed while solving the first $p$ right-hand sides with Bl-GMRES-DR(m,p,k).
 \item Apply one cycle of Bl-GMRES(m,p).
 \item Test the residual norms for convergence (can also test during the Bl-GMRES cycles).  If not satisfied, go back to step 2.
\end{enumerate} 
\vspace{.15in}

{\it Example 2.}  We test the matrix of Example 1 for 40 right-hand sides.  See Table 4.1 for the results.  The method Bl-G-DR(170,20,10) + Bl-G(160,20)-Proj(10) means that Block-GMRES-DR with block-size of 20, 10 approximate eigenvectors, and total subspaces of dimension 170 is applied to the first 20 right-hand sides.  Then for the next 20 right-hand sides, minres projection over the 10 approximate eigenvectors is alternated with Block-GMRES using block-size of 20 and subspaces of maximum dimension 160.  However, for this problem Block-GMRES-DR does not converge.  The dimension of the Krylov subspace generated for each right-hand side is just 8, which is not enough for this difficult problem.  With block-size of 5, the method Bl-G-DR(170,5,10) + Bl-G(160,5)-Proj(10) uses many more flops than the non-block GMRES-Proj approach, but it does use less matrix-vector products.  Block-QMR uses even fewer matrix-vector products, and it can potentially take advantage of applying 40 matrix-vector products simultaneously.  Block-QMR builds a very large subspace that eventually contains approximations to many eigenvectors, thus giving this rapid convergence.  If one is only interested exclusively in the number of matrix-vector products, Bl-GMRES-DR can actually be the winner.  Only 2400 matrix-vector products are needed for Bl-GMRES-DR(1220,40,20).

\begin{table}
\caption{Comparison of block-GMRES-Proj with other methods for 40 right-hand sides} 

\begin{center} \footnotesize
\begin{tabular}{|c|c|c|c|}  \hline\hline
       & matrix-vector products  &  flops     \\  
\hline \hline
GMRES-DR(25,10) + GMRES(15)-Proj(10)       &  5151   &   6.1   \\ \hline
Bl-G-DR(170,20,10) + Bl-G(160,20)-Proj(10)   &   -     &  -      \\ \hline
Bl-G-DR(170,10,10) + Bl-G(160,10)-Proj(10)   &  4960   &  170.4  \\ \hline
Bl-G-DR(170,5,10) + Bl-G(160,5)-Proj(10)     &  4169   &  116.6  \\ \hline
Bl-G-DR(60,5,10) + Bl-G(50,5)-Proj(10)       &  5900   &   30.9  \\ \hline
Bl-QMR                                     &  3298   &  117.2  \\ \hline

\hline\hline 

\end{tabular} 
\end{center} 
\end{table}

\section{Application to Quantum Chromodynamics}

\subsection{GMRES-Proj for a QCD problem}

We demonstrate the GMRES-Proj method for an application from particle physics.  In lattice quantum chromodynamics (QCD), very large systems of linear equations arise that have complex non-Hermitian matrices.  For such matrices, we need to change transpose to Hermitian transpose in the algorithms.  Often there are multiple right-hand sides for each QCD matrix.  However, block methods are not typically used.  The matrix-vector product is moderately expensive (it can be implemented for a cost equivalent to 72 non-zeros per row~\cite{FrMe} even though there would actually be about three times as many non-zeros in the matrix if it was formed).  The orthogonalization costs are significant enough to discourage block methods.  Therefore it would be very useful to improve convergence of the main methods used  for QCD problems such as restarted GMRES and BiCGStab.  Our approach to deflation of multiple right-hand sides is new; however, deflation in the context of lattice problems was originally considered in~\cite{dF}.  See~\cite{EdHeNa,DoLeLiZh,NeEiLiNeSc} for other approaches.  

{\it Example 3.}
We look at a typical Wilson-Dirac matrix from QCD.  It has even-odd preconditioning~\cite{Qcdconf}, and the size is 248,832 by 248,832.  The value of $\kappa$ is 0.159, which is approximately $\kappa_{critical}$, so the leftmost eigenvalues are near the imaginary axis.  The right-hand sides are unit vectors associated with particular space-time, Dirac and color coordinates.  The first right-hand side is solved with GMRES-DR(50,30) to three different residual tolerances.  Then for the second right-hand side, GMRES-Proj uses 30 approximate eigenvectors for the projection in between cycles of GMRES(20).  See Figure 4.1 for the results.  Solving the first right-hand side to one of the more demanding tolerances ($10^{-10}$ or $10^{-14}$) pays off.  GMRES(20)-Proj(30) can converge in less than one-tenth of the iterations needed for GMRES(20).   

\begin{figure}
\vspace{.10in}
\begin{center}
\includegraphics[scale=0.5]{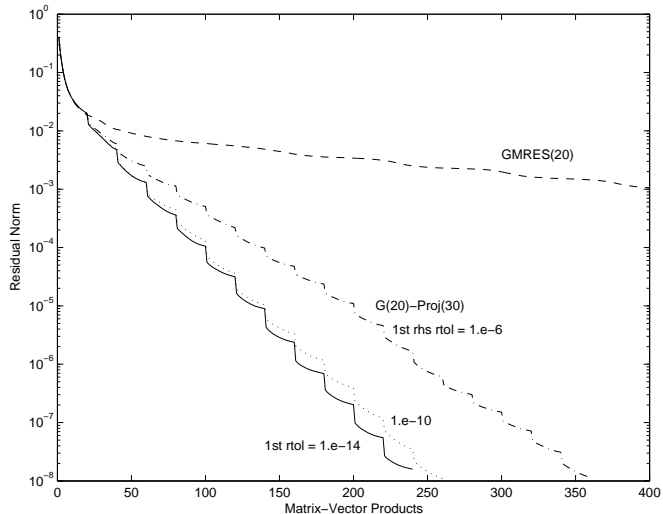}
\end{center}
\vspace{.10in}
\caption{Solution of second RHS for large QCD matrix.}
\end{figure}

Figure 5.2 shows convergence with different frequencies of projection for GMRES(20)-Proj(30) with the first right-hand side solved to $rtol = 10^{-10}$ .  Projecting in between every cycle turns out a little better (for a different QCD matrix in~\cite{Qcdconf}, projecting every third cycle was as effective as every cycle). 

\begin{figure}
\vspace{.10in}
\begin{center}
\includegraphics[scale=0.5]{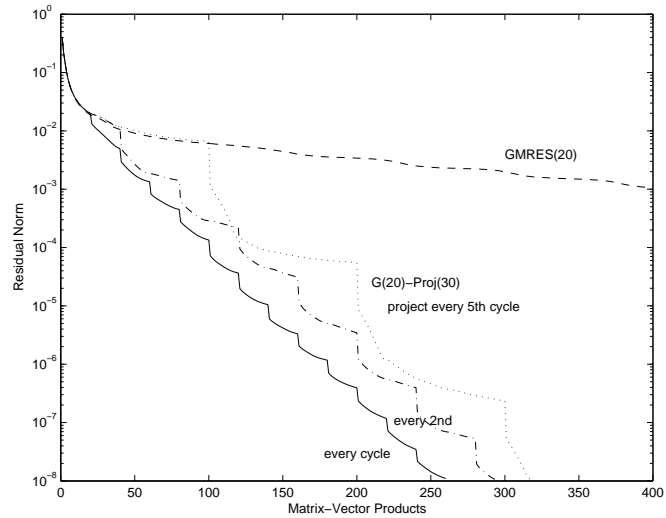}
\end{center}
\vspace{.10in}
\caption{Different frequencies of projection for the QCD matrix.}
\end{figure}

Figure 5.3 shows the harmonic Ritz values generated by GMRES-DR(50,30) after 50, 550 and 890 matrix-vector products.  The case of 550 corresponds to $rtol = 10^{-10}$.  Figure 3.5 has a blowup of the portion of that graph near the origin.  After 550 iterations, the small approximate eigenvalues are settling in near to where they are after 890.  These graphs show why deflating eigenvalues is so effective for this problem.  The origin is halfway surrounded by eigenvalues, until the smallest ones are deflated.

\begin{figure}
\vspace{.10in}
\begin{center}
\includegraphics[scale=0.5]{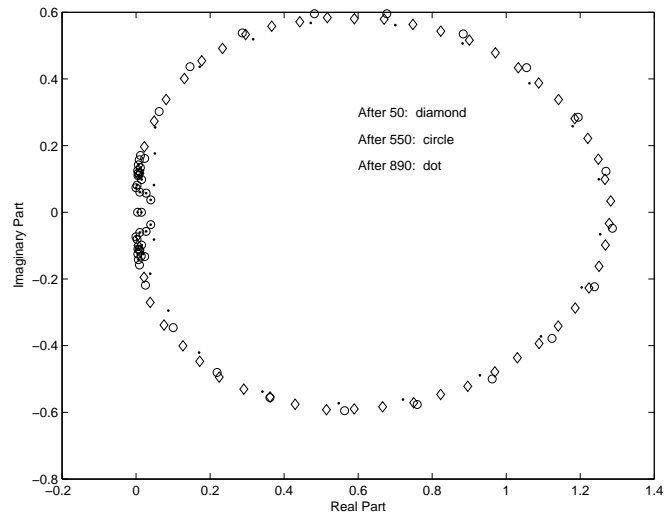}
\end{center}
\vspace{.10in}
\caption{Harmonic Ritz values from GMRES-DR with the QCD matrix.}
\end{figure}

\begin{figure}
\vspace{.10in}
\begin{center}
\includegraphics[scale=0.5]{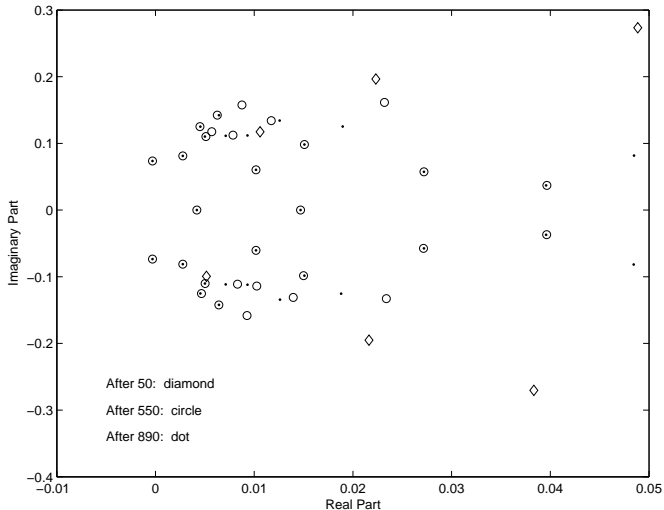}
\end{center}
\vspace{.10in}
\caption{Closeup of harmonic Ritz values near the origin.}
\end{figure}

Next an example is given that has comparison with perhaps the most popular approach to deflating eigenvalues in QCD, which is to compute eigenvectors in a separate routine, then use them in any of several ways to deflate.

{\it Example 4.}  
Here we use a small QCD matrix.  It is complex with dimension 1536.  The leftmost two eigenvalues have small negative real parts, so the $\kappa$ value has gone just past $\kappa_{critical}$.  We compare computing eigenvectors with GMRES-DR and the MATLAB routine Eigs.  GMRES-DR(30,16) solves the first system to residual norm $10^{-8}$, then GMRES(14)-Proj(16) is applied to the other right-hand sides.   For the other approach, Eigs is called to compute 16 eigenvectors.  This implements implicitly restarted Arnoldi~\cite{So}.  We do not allow the matrix to be factored.  After Eigs has computed the eigenvectors nearly exactly, the deflation can be done in various ways, but here we again use GMRES-Proj.  The non-deflated method GMRES(16) is also compared.  The results are in Table 5.1 with the number of matrix-vector products given for each approach.  For this relatively tiny matrix, the small eigenvalues are not as much of a problem as in the previous example, however the deflated methods still use less matrix-vector products than regular GMRES.  Computing the eigenvectors with GMRES-DR is more efficient than computing them in a separate routine.  GMRES-DR takes 84 matrix-vector products compared to 384 for Eigs, and it also solves the first system of linear equations at the same time.  Eigs does generate more accurate eigenvectors.  However, for the second and subsequent right-hand sides, the convergence for GMRES-Proj is almost identical regardless of which routine found the eigenvectors.

\begin{table}
\caption{Matrix-vector products for small QCD matrix with 12 right-hand sides} 

\begin{center} \footnotesize
\begin{tabular}{|c|c|}  \hline\hline
GMRES(14)         					& 1204   	\\ \hline
GMRES-DR(30,16) + GMRES(14)-Proj(16)        	& 720   	\\ \hline
Eigs + GMRES(14)-Proj(16)         		    	& 1069   	\\ \hline 

\hline\hline 

\end{tabular} 
\end{center} 
\end{table} 

\section{Conclusion}

In this paper, we have shown that deflating eigenvalues can be very helpful for solving systems with multiple right-hand sides.  The first right-hand side is solved with the deflated GMRES method GMRES-DR.  This method develops eigenvector information that is used for all subsequent right-hand sides.  Therefore there is no requirement that the right-hand sides all be available simultaneously.  Also, since the needed eigenvector information is available from the beginning for the subsequent right-hand sides, the convergence can be much faster, particularly for tough problems with small eigenvalues.  The approach in GMRES-Proj of projecting in between cycles of GMRES is very efficient.  For the case of related right-hand sides, there is a simple, but especially effective approach.

Block methods are a competing approach to eigenvalue deflation methods.  However, we have also looked at combining block methods with deflation.

In lattice QCD physics, very large systems with multiple right-hand sides need to be solved.  In one example, GMRES-Proj is an order of magnitude better that regular restarted GMRES.  Deflated versions of BiCGStab for QCD problems will be studied in the future.  

Future research will focus on deflating eigenvalues for QCD problems which not only have multiple right-hand sides, but have multiple shifts of the matrix for each right-hand side.  The goal is to solve all the shifted systems for approximately the same cost as solving one.  It would also be worthwhile to investigate deflation for other QCD problems such as twisted mass and overlap fermions.

\section*{Acknowledgments} The first author wishes to thank Andreas Frommer for helpful discussions. 

\bibliography{morgan}

\end{document}